\documentclass[conference]{IEEEtran}
\IEEEoverridecommandlockouts
\usepackage{cite}
\usepackage{amsmath,amssymb,amsfonts}
\usepackage{algorithm}
\usepackage{algorithmic}
\usepackage{setspace}
\usepackage{graphicx}
\usepackage{subcaption}
\usepackage{textcomp}
\usepackage{xcolor}
\usepackage{multirow}
\usepackage{hyperref} 
\usepackage[flushleft]{threeparttable}
\usepackage{balance} 
\usepackage{verbatim} 
\def\BibTeX{{\rm B\kern-.05em{\sc i\kern-.025em b}\kern-.08em
    T\kern-.1667em\lower.7ex\hbox{E}\kern-.125emX}}

\usepackage{tikz}
\usetikzlibrary{matrix,decorations.pathreplacing,positioning}

\begin{document}

\title{An Iterated Hybrid Fast Parallel FIR Filter \\
}
\author{
\IEEEauthorblockN{Keshab K. Parhi,~\IEEEmembership{Life Fellow,~IEEE}}
\IEEEauthorblockA{Department of Electrical and Computer Engineering \\
University of Minnesota, Minneapolis, MN, USA \\
Email: parhi@umn.edu}
}
\maketitle

\begin{abstract}
This paper revisits the design and optimization of parallel fast finite impulse response (FIR) filters using polyphase decomposition and iterated fast FIR algorithms (FFAs). Parallel FIR filtering enhances computational efficiency and throughput in digital signal processing (DSP) applications by enabling the simultaneous processing of multiple input samples. We revisit a prior approach to design of fast parallel filter architectures by using the iterated FFA approach where the same primitive filter, such as 2-parallel, is iterated to design the fast parallel filter. In this paper, we present yet another novel iterated fast parallel FIR filter, referred to as the \textit{fast hybrid} filter. The hybrid filter iterates a transposed 2-parallel fast FIR filter in all the inner layers and a direct-form 2-parallel fast FIR filter in the outermost layer, resulting in reduced hardware complexity. Such an iterated hybrid approach has not been presented before. We show that the hybrid fast parallel filters require less number of additions compared to prior approaches. 

\end{abstract}
\section{Introduction}
\label{sec:intro}

\begin{figure*}
\centering
\begin{subfigure}{0.48\textwidth}
    \includegraphics[width=\textwidth]{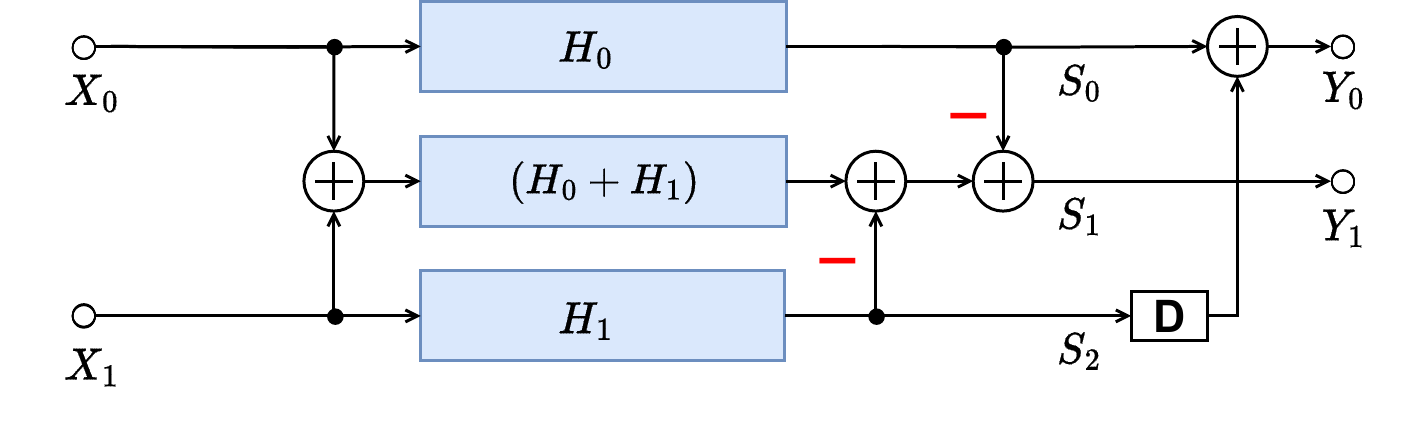}
    \caption{Low-complexity 2-parallel fast filter (direct plus form).}
    \label{fig:2pdp}
\end{subfigure}
\hfill
\begin{subfigure}{0.48\textwidth}
    \includegraphics[width=\textwidth]{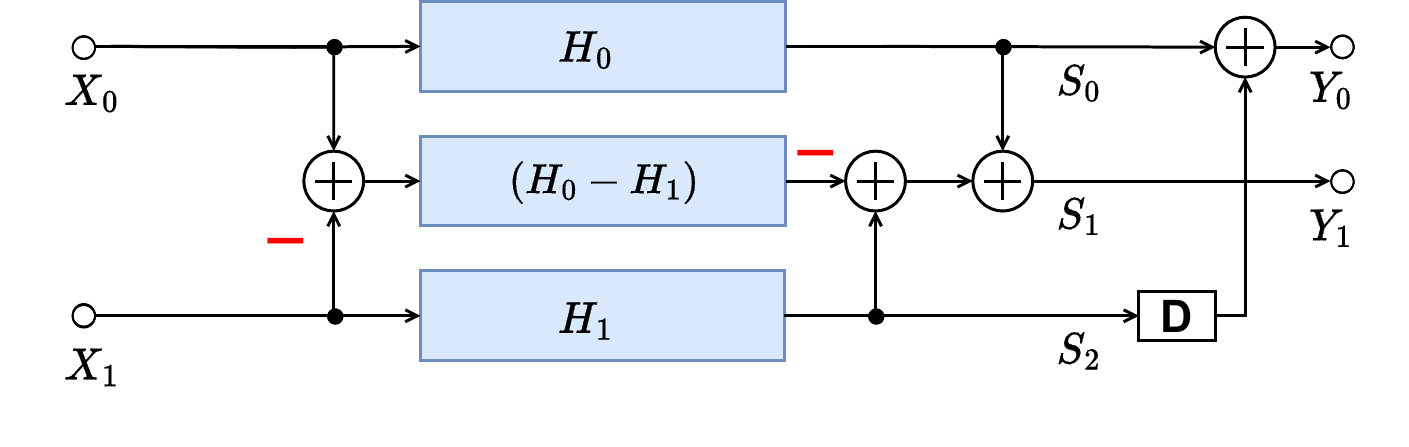}
    \caption{Low-complexity 2-parallel fast filter (direct minus form).}
    \label{fig:2pdm}
\end{subfigure}
\hfill
\par\bigskip 
\begin{subfigure}{0.48\textwidth}
    \includegraphics[width=\textwidth]{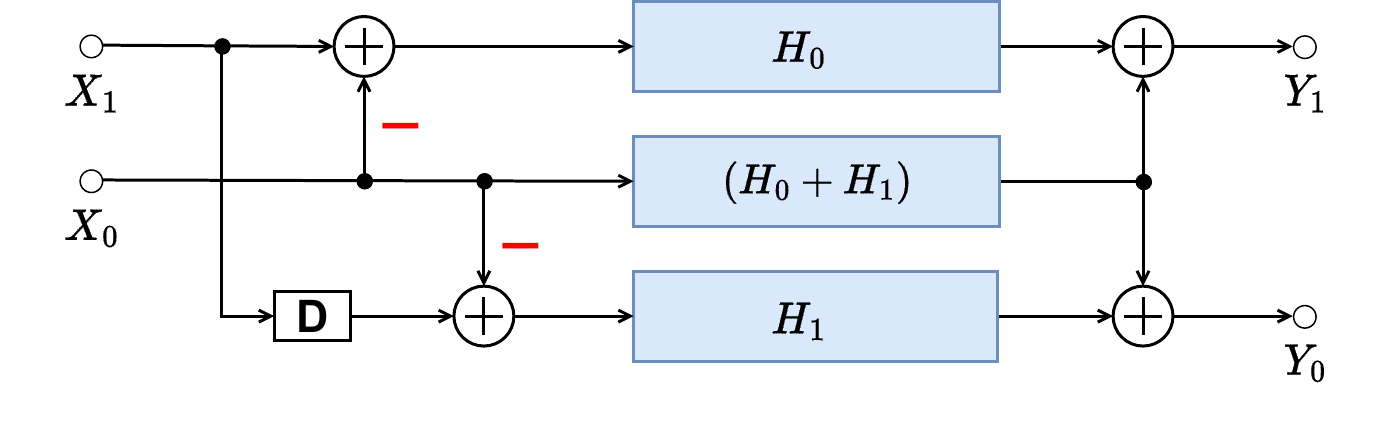}
    \caption{Low-complexity 2-parallel fast filter (transposed plus form).}
    \label{fig:2ptp}
\end{subfigure}
\hfill
\begin{subfigure}{0.48\textwidth}
    \includegraphics[width=\textwidth]{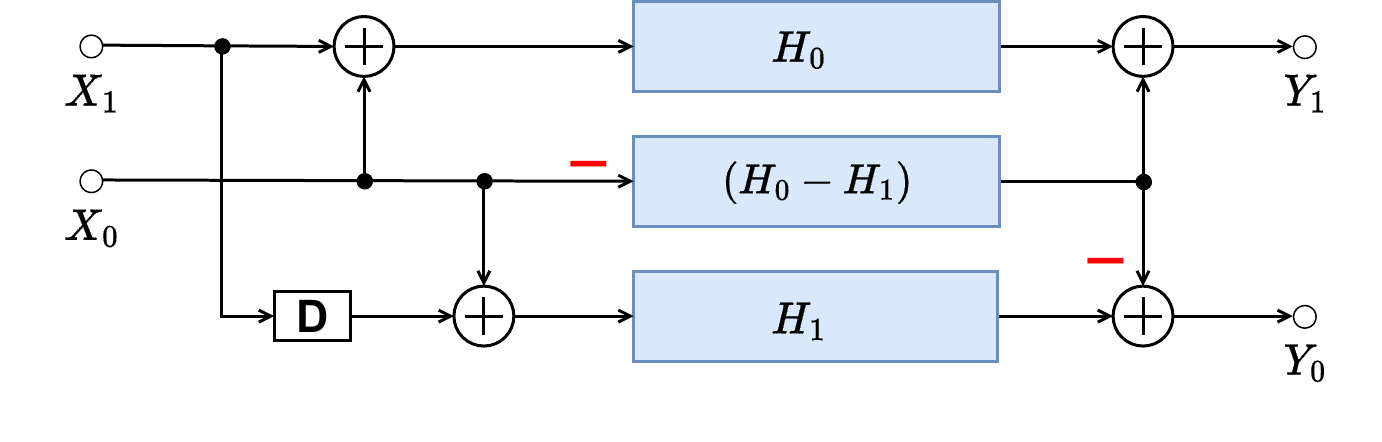}
    \caption{Low-complexity 2-parallel fast filter (transposed minus form).}
    \label{fig:2ptm}
\end{subfigure}
        
\caption{Low-complexity 2-parallel fast filters.}
\label{fig:2p}
\end{figure*}


Finite impulse response (FIR) filters are fundamental building blocks in digital signal processing (DSP) systems, widely used in applications ranging from video and image processing to wireless communications. High-performance DSP applications such as optical transmission often require FIR filters to be operated at high sample speeds with low area and low power consumption. Parallel processing techniques, based on polyphase decomposition, can be employed to improve computational efficiency \cite{parhi2007vlsi}. Fast parallel FIR flters can be derived from iterative applications of fast convolution structures \cite{cheng2004} and by fast filter algorithms \cite{parhi2007vlsi}. This paper reviews the design of iterated fast parallel filters using fast filter algorithms (FFAs), and presents a yet another novel parallel filter structure, referred as the
\textit{fast hybrid} parallel FIR filter. While the fast convolution approach \cite{cheng2004} is not discussed in the paper, the comparison section includes the addition and delay comparisons of this approach.

Consider the formulation of parallel FIR filters using polyphase decomposition, a technique widely used in multirate signal processing \cite{vaidyanathan2006multirate}. An \(N\)-tap FIR filter can be expressed in the time domain as:
\begin{align}
    y(n) &= h(n) * x(n) \nonumber \\
    &= \sum_{i=0}^{N-1} h(i)x(n - i), n = 0,1,2,\dots,\infty,
\end{align}
where \( x(n) \) is an infinite-length input sequence, and the sequence \( h(n) \) contains FIR filter coefficients of length-\(N\).  Similarly, in the \(Z\)-domain, this can be expressed as:
\begin{equation}
    Y(z) = H(z)X(z) 
\end{equation}
where $X(z)$ is given by:
\begin{equation}
    X(z) = x(0) + x(1)z^{-1} + x(2)z^{-2} + x(3)z^{-3} + \dots
\end{equation}
The input sequence \( \{ x(0), x(1), x(2), x(3), \dots \} \) can be decomposed into even and odd samples as:
\begin{align}
    X(z) &= x(0) + x(2)z^{-2} + x(4)z^{-4} + \dots \nonumber \\
    &+ z^{-1}[x(1) + x(3)z^{-2} + x(5)z^{-4} + \dots]
\end{align}
leading to the 2-phase polyphase decomposition:
\begin{equation}
    X(z) = X_0(z^2) + z^{-1}X_1(z^2),
\end{equation}
where \( X_0(z^2) \) and \( X_1(z^2) \) are the \(Z\)-transforms of \( x(2k) \) and \( x(2k+1) \), respectively, for \( 0 \leq k < \infty \).

Similarly, the Z-transform of the length-\(N\) filter, \(H(z)\), can be decomposed as:
\begin{equation}
    H(z) = H_0(z^2) + z^{-1}H_1(z^2),
\end{equation}
where \( H_0(z^2) \) and \( H_1(z^2) \) are of length-\(N/2\), and are referred to as the even and odd subfilters, respectively. For example, the even and odd subfilters of a 6-tap FIR filter \( H(z) = h(0) + h(1)z^{-1} + h(2)z^{-2} + h(3)z^{-3} + h(4)z^{-4} + h(5)z^{-5} \) are:
\begin{align}
    H_0(z^2) &= h(0) + h(2)z^{-2} + h(4)z^{-4}, \\
    H_1(z^2) &= h(1) + h(3)z^{-2} + h(5)z^{-4}.
\end{align}

The even output sequence, \( y(2k) \), and the odd output sequence, \( y(2k+1) \), (for \( 0 \leq k < \infty \)) can be expressed as:
\begin{equation}
    Y(z) = Y_0(z^2) + z^{-1}Y_1(z^2),
\end{equation}
where:
\begin{align}
    Y_0(z^2) &= X_0(z^2)H_0(z^2) + z^{-2}X_1(z^2)H_1(z^2), \nonumber \\
    Y_1(z^2) &= X_0(z^2)H_1(z^2) + X_1(z^2)H_0(z^2). \label{eqn:2pfir}
\end{align}





Since the work of Winograd \cite{winograd1980arithmetic}, it has been established that two polynomials of degree \( L-1 \) can be multiplied using only \( (2L - 1) \) product terms \cite{blahut2010fast}. 
Fast FIR algorithms (FFAs) \cite{mou1987fast, mou1991short}, leverage this approach to develop lower-complexity parallel filtering structures \cite{parker1997low}. By applying this method, an \( L \)-parallel FIR filter can be implemented with approximately \( (2L - 1) \) filtering operations of length \( N/L \). Consequently, the resulting parallel filtering structure requires only \( (2N - N/L) \) multiplications, significantly reducing multiplication complexity, at the expense of addition complexity.

Reducing the number of multiplications and additions in parallel FIR filters has implications in other problems such as parallel modular multiplication \cite{TanPQCTC,IngridJSSC,fan2010overlap}, and fast pointwise multiplication for matrix-vector polynomial modular multiplication for post-quantum cryptography \cite{TanICCAD,TanISCAS2024,zhou2018preprocess,xing2021compact,zhu2021ntt}, as fast parallel filters can form the basis for equivalent structures for these operations \cite{ParhiAsilomar2025}.

The main contribution of this paper is design of fast hybrid parallel filters where different primitive structures are iterated at different levels. Specifically, the fast hybrid filter is designed using a direct-form fast filter as the outer layer and the transpose-form fast filter as the inner layers of the iterations; thus the name \textit{hybrid}.

\begin{figure*}
\centering
\begin{subfigure}{0.46\textwidth}
    \includegraphics[width=\textwidth]{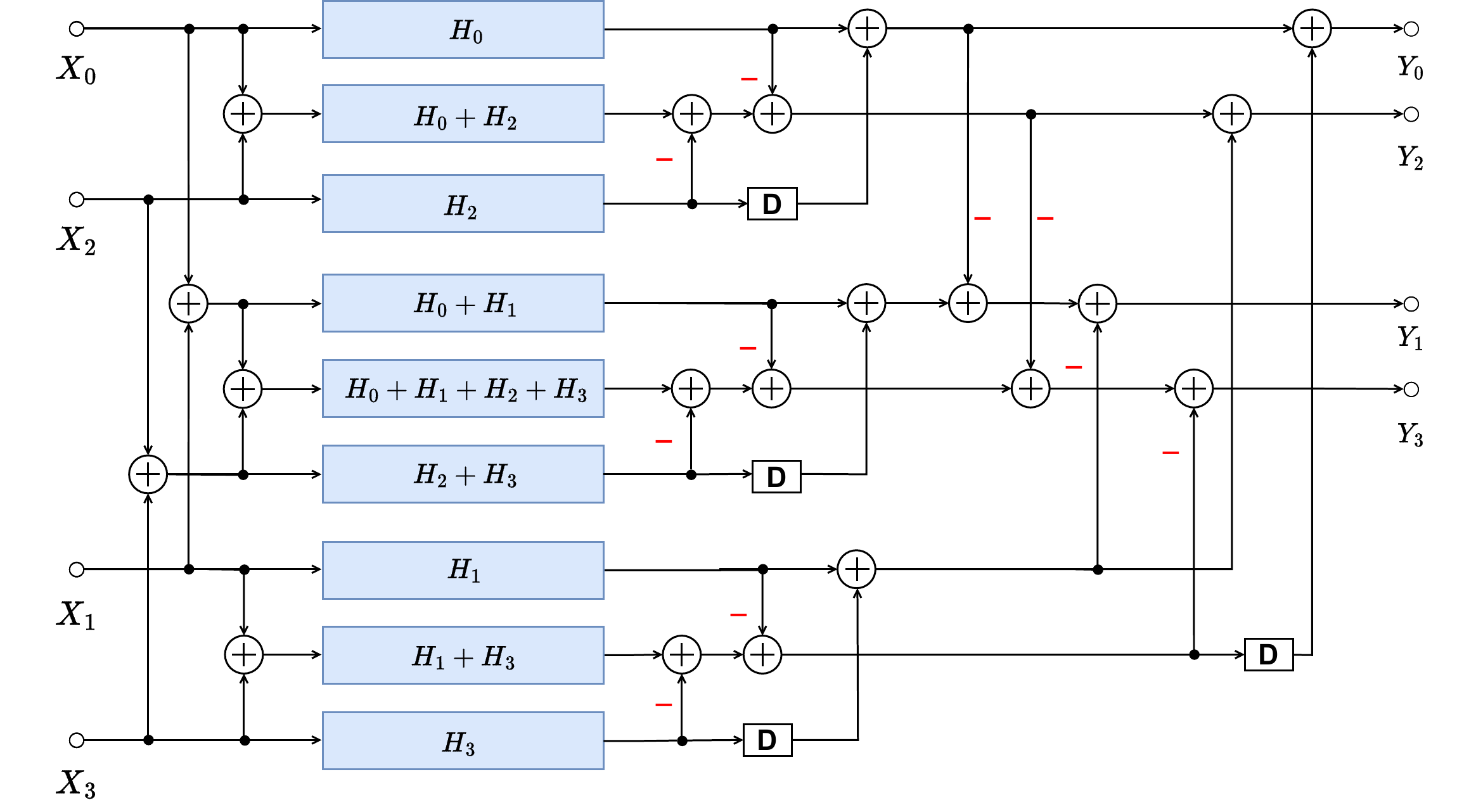}
    \caption{Direct form.}
    \label{fig:4pd}
\end{subfigure}
\hfill
\begin{subfigure}{0.48\textwidth}
    \includegraphics[width=\textwidth]{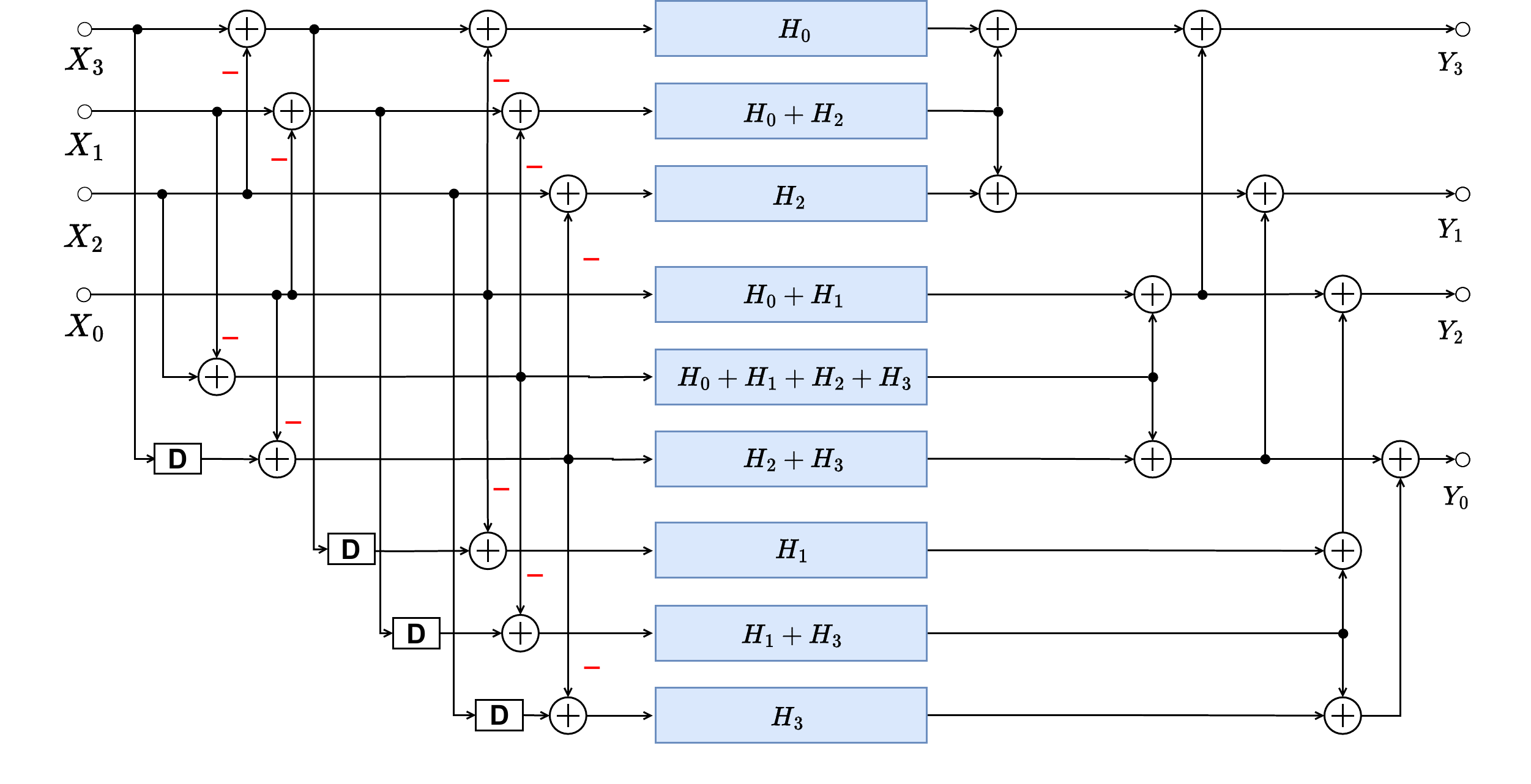}
    \caption{Transposed form.}
    \label{fig:4pt}
\end{subfigure}
        
\caption{4-parallel fast filter.}
\label{fig:4p}
\end{figure*}

\begin{figure*}
\centering
\begin{subfigure}{0.45\textwidth}
    \includegraphics[width=\textwidth]{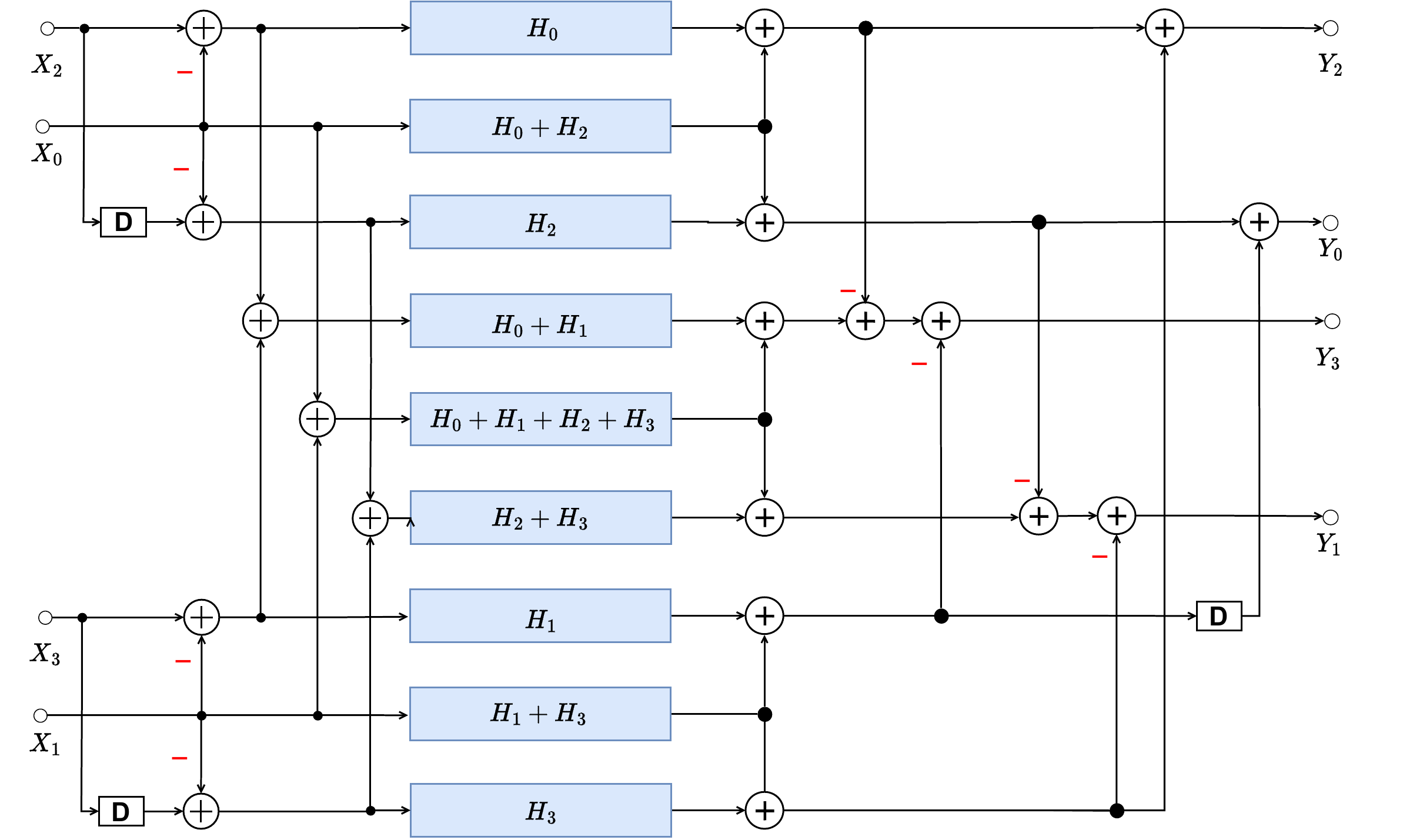}
    \caption{Direct form.}
    \label{fig:4phd}
\end{subfigure}
\hfill
\begin{subfigure}{0.5\textwidth}
    \includegraphics[width=\textwidth]{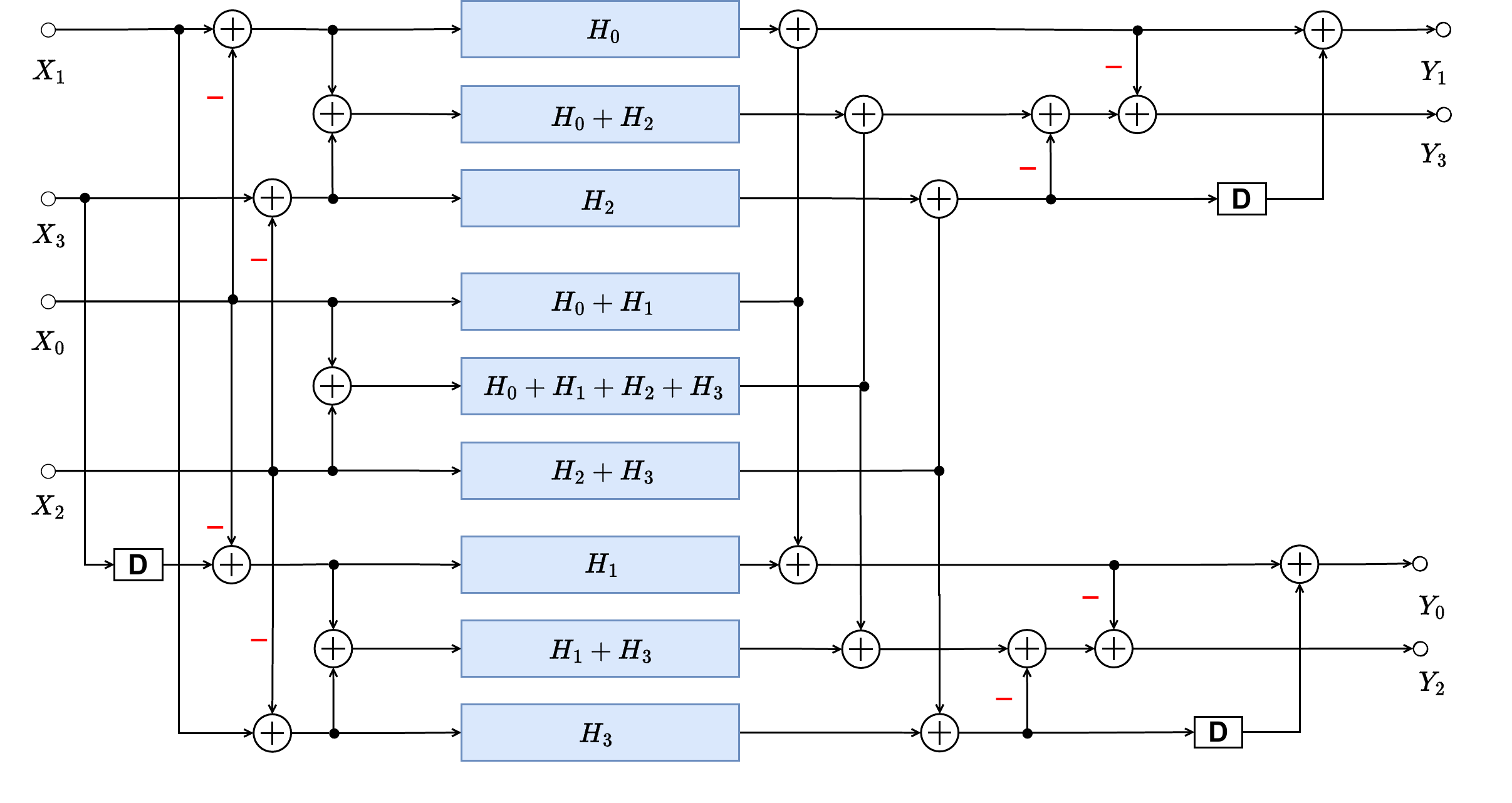}
    \caption{Transposed form.}
    \label{fig:4pht}
\end{subfigure}
        
\caption{4-parallel hybrid fast filter.}
\label{fig:4ph}
\end{figure*}

\begin{figure*}[htb!]
\centering
\resizebox{0.8\textwidth}{!}{
\includegraphics{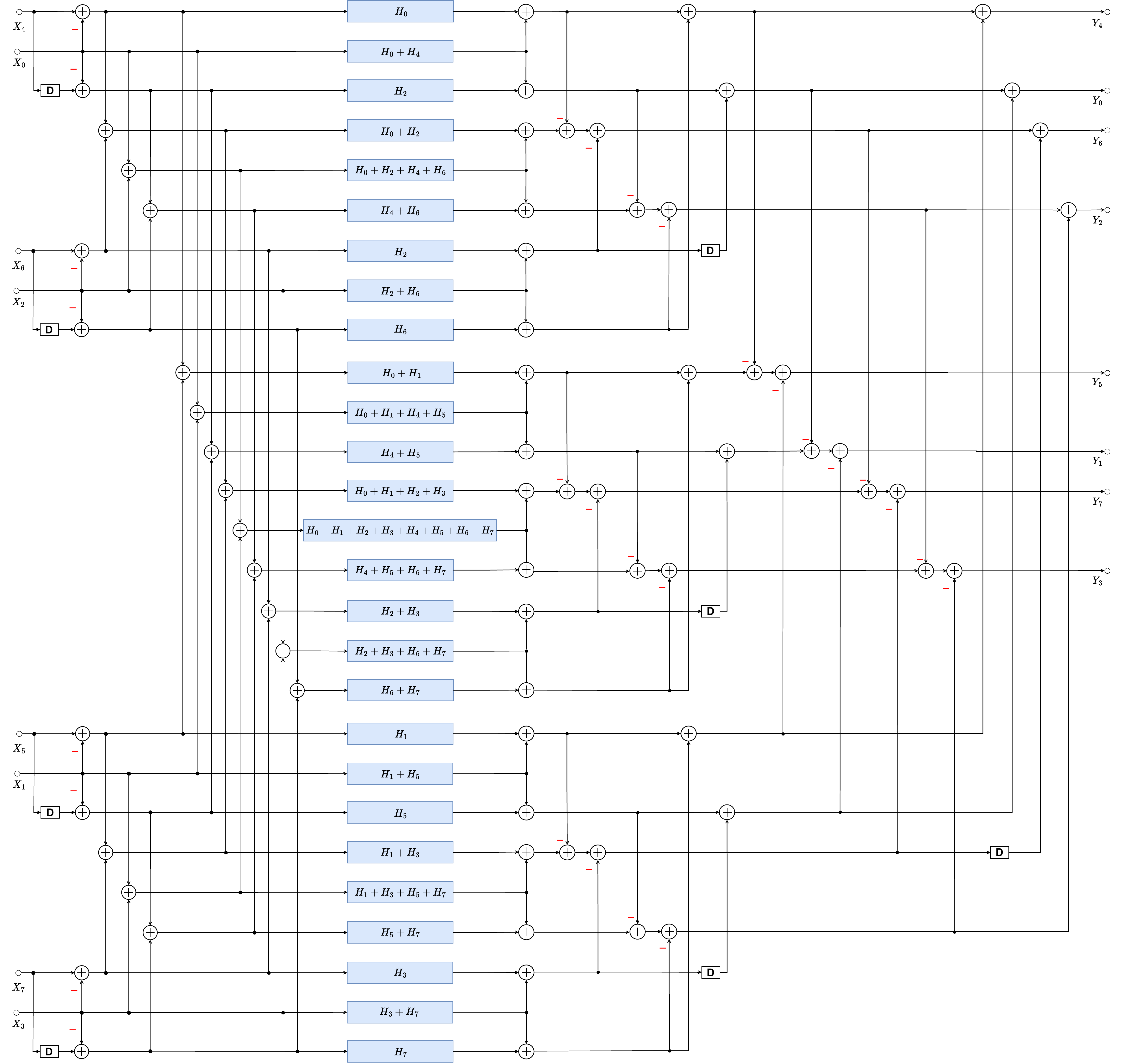}}
\caption{8-parallel fast iterative FIR hybrid filter.} 
\label{fig:8ph}
\end{figure*}

\section{Power-of-2 Parallel Fast filtering structure by Iterated FFA}
A 2-parallel FFA is described by:
\begin{equation}
    Y_0 = H_0 X_0 + z^{-2} H_1 X_1
\end{equation}
\begin{equation}
    Y_1 = (H_0 + H_1)(X_0 + X_1) - H_0 X_0 - H_1 X_1
\end{equation}
and is shown in Fig. \ref{fig:2pdp}. Another 2-parallel FFA using the minus form is shown in Fig. \ref{fig:2pdm}. Figs. \ref{fig:2ptp} and \ref{fig:2ptm}, respectively, describe their transpose forms. It is well known that transpose forms of parallel FIR filters with flipped inputs and outputs describe equivalent parallel FIR filters \cite{parhi2007vlsi}.

We can extend the FFA concept to the power-of-2 parallel fast filters \cite{parhi2007vlsi}. Considering the design of a 4-parallel filter, we have

\begin{align}
    Y = &Y_0 + z^{-1} Y_1 + z^{-2} Y_2 + z^{-3} Y_3 \nonumber \\
      = &(X_0 + z^{-1} X_1 + z^{-2} X_2 + z^{-3} X_3) \nonumber \\
      &(H_0 + z^{-1} H_1 + z^{-2} H_2 + z^{-3} H_3). \label{eqn:4p}
\end{align}

The reduced-complexity 4-parallel fast FIR filtering structure is obtained by first applying the 2-parallel FFA to Eqn. (\ref{eqn:4p}) and then applying the FFA a second time. From Eqn. (\ref{eqn:4p}), we have:

\begin{align}
    Y &= (X_0' + z^{-1} X_1')(H_0' + z^{-1} H_1'), \label{eqn:4pFFA1}
\end{align}
where

\begin{align*}
    X_0' &= X_0 + z^{-2} X_2, & X_1' &= X_1 + z^{-2} X_3, \\
    H_0' &= H_0 + z^{-2} H_2, & H_1' &= H_1 + z^{-2} H_3.
\end{align*}

Eqn. (\ref{eqn:4pFFA1}) can be rewritten as:
\begin{align}
    Y &= X_0' H_0' + z^{-2} X_1' H_1'. \nonumber \\
    &+ z^{-1} \big[(X_0' + X_1')(H_0' + H_1') - X_0' H_0' - X_1' H_1'\big]
\end{align}

The same 2-parallel FFA is then applied a second time to each of the filtering operations \( X_0' H_0' \), \( X_1' H_1' \), and \( (X_0' + X_1')(H_0' + H_1') \).

The second application results in the 4-parallel fast filtering structure shown in Fig. \ref{fig:4pd} \cite{parhi2007vlsi}. This structure is obtained by iterating the 2-parallel FFA in Fig. \ref{fig:2pdp} twice. It requires 4 delay elements, and 20 additions. Transposing this structure leads to another 4-parallel FIR filter in Fig. \ref{fig:4pt}.

The 4-parallel fast filter in Fig. \ref{fig:4pd} can also be written in matrix form:
{
\begin{align}
&\mathbf{P_4}\mathbf{Y_{4}}= \mathbf{D_{4}} (\mathbf{Q_2}\otimes\mathbf{I_2}) (\mathbf{I_{3}}\otimes \mathbf{D_{24}Q_2}) \nonumber \\
&~ \text{diag}\left( ( \mathbf{P_2}\otimes \mathbf{P_2}) \mathbf{P_4}\mathbf{H_{4}}\right) ( \mathbf{P_2}\otimes \mathbf{P_2}) \mathbf{P_4} \mathbf{X_{4}}, 
\end{align}
}
where
{\footnotesize
\begin{equation*}
    \mathbf{P_4} = 
    \begin{bmatrix}
        1 & 0 & 0 & 0 \\
        0 & 0 & 1 & 0 \\
        0 & 1 & 0 & 0 \\
        0 & 0 & 0 & 1 \\
    \end{bmatrix},
    \mathbf{D_{4}} = 
    \begin{bmatrix}
        1 & 0 & 0 & 0 & 0 & z^{-4} \\
        0 & 1 & 0 & 0 & 1 & 0 \\
        0 & 0 & 1 & 0 & 0 & 0 \\
        0 & 0 & 0 & 1 & 0 & 0 \\
    \end{bmatrix},
\end{equation*}
\begin{equation*}
    \mathbf{D_{24}} = 
    \begin{bmatrix}
        1 & 0 & z^{-4} \\
        0 & 1 & 0
    \end{bmatrix}.
\end{equation*}
\begin{equation*}
    \mathbf{Q_2} = 
    \begin{bmatrix}
        1 & 0 & 0 \\
        -1 & 1 & -1 \\
        0 & 0 & 1
    \end{bmatrix},
    \mathbf{P_2} = 
    \begin{bmatrix}
        1 & 0 \\
        1 & 1 \\
        0 & 1
    \end{bmatrix}.
    \mathbf{X_{4}} = 
    \begin{bmatrix}
        X_0 \\
        X_1 \\
        X_2 \\
        X_3
    \end{bmatrix}.
\end{equation*}
}
The notation $\mathbf{I_N}$ represents an $N \times N$ identity matrix. The number of additions for a $2^n$-parallel filter derived by iterated FFA is given by:
\begin{equation}
    A_n = 4(3^n-2^n)
\end{equation}
The number of delay elements is given by $(3^n -1)/2$.



\section{Power-of-2 Parallel Fast Iterated FIR Hybrid Filter}

Recall that in Fig. \ref{fig:4pd}, both the inner and outer layers consist of the 2-parallel fast filter shown in Fig. \ref{fig:2pdp}. Suppose we construct a 4-parallel fast filter using the 2-parallel fast filter in transposed plus form from Fig. \ref{fig:2ptp} as the inner layer and the 2-parallel fast filter in direct form from Fig. \ref{fig:2pdp} as the outer layer. In that case, we obtain the 4-parallel fast filter structure illustrated in Fig. \ref{fig:4phd}, after using substructure sharing at the input side. We refer to this structure as the fast {\em hybrid} filter since the inner layer is in transposed form and the outer layer is in direct form. It requires 3 delay elements, and only 19 additions. This structure saves one delay element and one addition, compared to a traditional iterated FFA structure. A transpose form of the hybrid structure is shown in \ref{fig:4pht}. 

This hybrid fast filter can also be written in matrix form:
{ 
\begin{align}
    &\mathbf{Y_{4p}}=\mathbf{D'_{4}}(\mathbf{Q_{2}}\otimes \mathbf{P_{2}^{T}}) \nonumber \\
    &~ \text{diag}\left(( \mathbf{P_2}\otimes \mathbf{P_2}) \mathbf{P_4}\mathbf{H_{4}}\right)~ (\mathbf{P_2}\otimes \mathbf{Q_2^T D_{24}^T})\mathbf{X_{4p}},
\end{align}
}

where
{\footnotesize	
\begin{equation*}
    \mathbf{Y_{4p}} = 
    \begin{bmatrix}
        Y_2 \\
        Y_0 \\
        Y_3 \\
        Y_1
    \end{bmatrix},
    \mathbf{D'_{4}} = 
    \begin{bmatrix}
        1 & 0 & 0 & 0 & 1 & 0 \\
        0 & 1 & 0 & 0 & 0 & z^{-4} \\
        0 & 0 & 1 & 0 & 0 & 0 \\
        0 & 0 & 0 & 1 & 0 & 0 \\
    \end{bmatrix},
    \mathbf{X_{4p}} = 
    \begin{bmatrix}
        X_2 \\
        X_0 \\
        X_3 \\
        X_1
    \end{bmatrix}.
\end{equation*}
}



Iterating the transpose form 2-parallel FFA as the inner layer twice and the direct-form 2-parallel FFA as the outer layer leads to the 8-parallel fast hybrid filter shown in Fig. \ref{fig:8ph}. This structure requires 8 delay elements and 71 additions. With substructure sharing in the post-processing part, the number of delay elements can be reduced to 7 while the number of additions increases to 73. Note that a traditional 8-parallel FFA requires 13 delay elements and 76 additions. One fundamental property of the fast hybrid filter is that the delay elements are distributed on both sides of the subfilters, whereas in iterated FFA, they are located on the output side only.

The number of additions for this structure is $(11/3)(3^n)-(7/2)(2^n)$ and the number of delay elements is given by $(2^n + 3^{n-1 } -1)/2$. Note that postprocessing optimization can be used to further reduce the number of delay elements, at the expense of additions.




\section{Comparisons}

In Table~\ref{tab:DesignCompAdd}, we compare the number of additions (A.) and delay elements (D. of $2^n$-parallel filters for $n = 4, 6$, and $8$, for the fast convolution approach \cite{cheng2004}, iterated FFA approach \cite{parhi2007vlsi}, and the proposed hybrid fast FFA approach. Note that the proposed fast hybrid filter requires the fewest number of additions among these structures.
\begin{table}[htbp!]
\setlength{\tabcolsep}{4pt}
\renewcommand{\arraystretch}{1.25}
\begin{center}
\begin{threeparttable}
\caption{The number of delay elements and additions in a $2^n$-parallel fast filter for $n=4, 6, 8$}
\label{tab:DesignCompAdd}
\begin{tabular}{|c|cc|cc|cc|}
\hline
\multirow{2}{*}{Design} & \multicolumn{2}{c|}{16-parallel} & \multicolumn{2}{c|}{64-parallel} & \multicolumn{2}{c|}{256-parallel} \\
 & A. & D. & A. & D. & A. & D.\\

\hline
Fast Convolution & 310 & 15 & 3,262 & 63 & 31,270 & 255\\
\hline
Iterated FFA & 260 & 40 & 2,660 & 364 & 25,220 & 3,280\\
\hline
Hybrid FFA & 241 & 21 & 2,449 & 153 & 23,161 & 1,221\\
\hline
\end{tabular}

	

\end{threeparttable}
\end{center}
\end{table}

\section{Conclusion}

We have shown that the proposed novel fast hybrid filter requires the fewest number of additions, compared to prior fast parallel filters. Future research will be directed towards analysis of finite word-length effects and hardware implementations, and investigation of applications of these systems for post-quantum cryptography and fast pointwise multiplication.

\section{Acknowledgment}
The author is grateful to Sin-Wei Chiu for his help in preparation of this paper. This paper was supported in part by the NSF under grant number CCF-2243053.

\bibliographystyle{IEEEtran}
\bibliography{refs}

\end{document}